\documentclass[aps,prl,twocolumn,english,balance,superscriptaddress,floats,showpacs,longbibliography]{revtex4-1}
\usepackage[T1]{fontenc}
\usepackage[latin9]{inputenc}
\usepackage{amsmath}
\usepackage{amssymb}
\usepackage{stmaryrd}
\usepackage{graphicx}
\usepackage{esint}
\usepackage{subfigure}
\usepackage{multirow}
\usepackage{wasysym}

\makeatletter

\newcommand{\beq}{\begin{equation}}
\newcommand{\eeq}{\end{equation}}
\newcommand{\bea}{\begin{eqnarray}}
\newcommand{\eea}{\end{eqnarray}}
\newcommand{\bwt}{\begin{widetext}}
\newcommand{\ewt}{\end{widetext}}
\@ifundefined{textcolor}{}
{%
 \definecolor{BLACK}{gray}{0}
 \definecolor{WHITE}{gray}{1}
 \definecolor{RED}{rgb}{1,0,0}
 \definecolor{GREEN}{rgb}{0,1,0}
 \definecolor{BLUE}{rgb}{0,0,1}
 \definecolor{CYAN}{cmyk}{1,0,0,0}
 \definecolor{MAGENTA}{cmyk}{0,1,0,0}
 \definecolor{YELLOW}{cmyk}{0,0,1,0}
}

\newcommand{\br}{\mathbf{r}}
\newcommand{\bR}{\mathbf{R}}
\newcommand{\bS}{\mathbf{S}}

\newcommand{\ii}{\mathrm{i}}

\newcommand{\fvec}[1]{\boldsymbol{#1}}

\newcommand{\half}{\frac{1}{2}}

\usepackage{babel}

\makeatother

\usepackage{babel}

\begin{document}

\title{Strong coupling phases of partially filled twisted bilayer graphene narrow bands}

\author{Jian Kang}
\email{jian.kang@fsu.edu}
\affiliation{National High Magnetic Field Laboratory, Tallahassee, Florida, 32304, USA}

\author{Oskar Vafek}
\email{vafek@magnet.fsu.edu}
\affiliation{National High Magnetic Field Laboratory, Tallahassee, Florida, 32304, USA}
\affiliation{Department of Physics,
Florida State University, Tallahassee, Florida 32306, USA}

\begin{abstract}
We identify states favored by Coulomb interactions projected onto the Wannier basis of the four narrow bands of the ``magic angle'' twisted bilayer graphene. At the filling of two electrons/holes per moire unit cell, such interactions favor an insulating $SU(4)$ ferromagnet. The kinetic terms select the ground state in which the two valleys with opposite spins are equally mixed, with vanishing magnetic moment per particle. We also find extended excited states, the gap to which decreases in magnetic field. An insulating stripe ferromagnetic phase is favored at one electron/hole per unit cell.
\end{abstract}

\maketitle

In addition to superconductivity, recent experiments on magic angle twisted bilayer graphene revealed insulating phases at carrier concentrations corresponding to partial occupation of the four narrow bands composite near the neutrality point~\cite{Pablo1,Pablo2,CoryAndrea}. Such correlated insulator phases seem to occur only when the bandwidth of the composite is reduced either by fine-tuning of the twist angle to the vicinity of the ``magic'' value $\sim1.1^\circ$ or by tuning the applied pressure at $\sim1.3^\circ$~\cite{Pablo1,Pablo2,CoryAndrea}. Importantly, the insulating states occur at {\it commensurate} (rational) fillings corresponding to $2$ electrons/holes per moire unit cell, with additional resistance peaks observed at fillings of $1$ hole/electron per unit cell and $3$ holes/electrons per unit cell~\cite{Pablo1,Pablo2,CoryAndrea}.
This observation is hard to reconcile with the notion that the insulation is due to Fermi surface nesting, or the van Hove singularities, reconstructed by electron-electron interactions, because such band structure features generically occur at {\it incommensurate} fillings.
Instead, the above observations suggest that the effective Coulomb interaction dominates the effective kinetic energy\cite{Pablo1,CoryAndrea}. The former is given by the projection of the Coulomb interaction onto the Hilbert space spanned by the narrow bands and is $\sim e^2/\epsilon \ell_m\sim 15meV$, where the moire period $\ell_m\sim 13nm$ and $\epsilon\approx 6$ is the dielectric constant of the encapsulating BN. The kinetic energy scale is given by the bandwidth. Although there is no direct measurement of the bandwidth, theoretical calculations routinely find it to be $\lesssim 10$meV~\cite{MacDonald,Pablo1,Koshino,KangVafek,AdrianPo,Tomanek,Fabrizio}.

Such considerations hint that, even if the physical system is ultimately in an intermediate coupling regime, a strong coupling approach may be more successful in capturing the nature of the correlated phases. In this approach the interaction-only Hamiltonian is minimized first, and the kinetic energy term is then treated as a perturbation\cite{Balents,Patrick,AdrianPo,Phillips,Kivelson,Sachdev,CKXu2,Kuroki2,Fernandes,Karrasch,PhillipsMI}.

\begin{figure}[htbp]
\centering
\includegraphics[width=0.6\columnwidth]{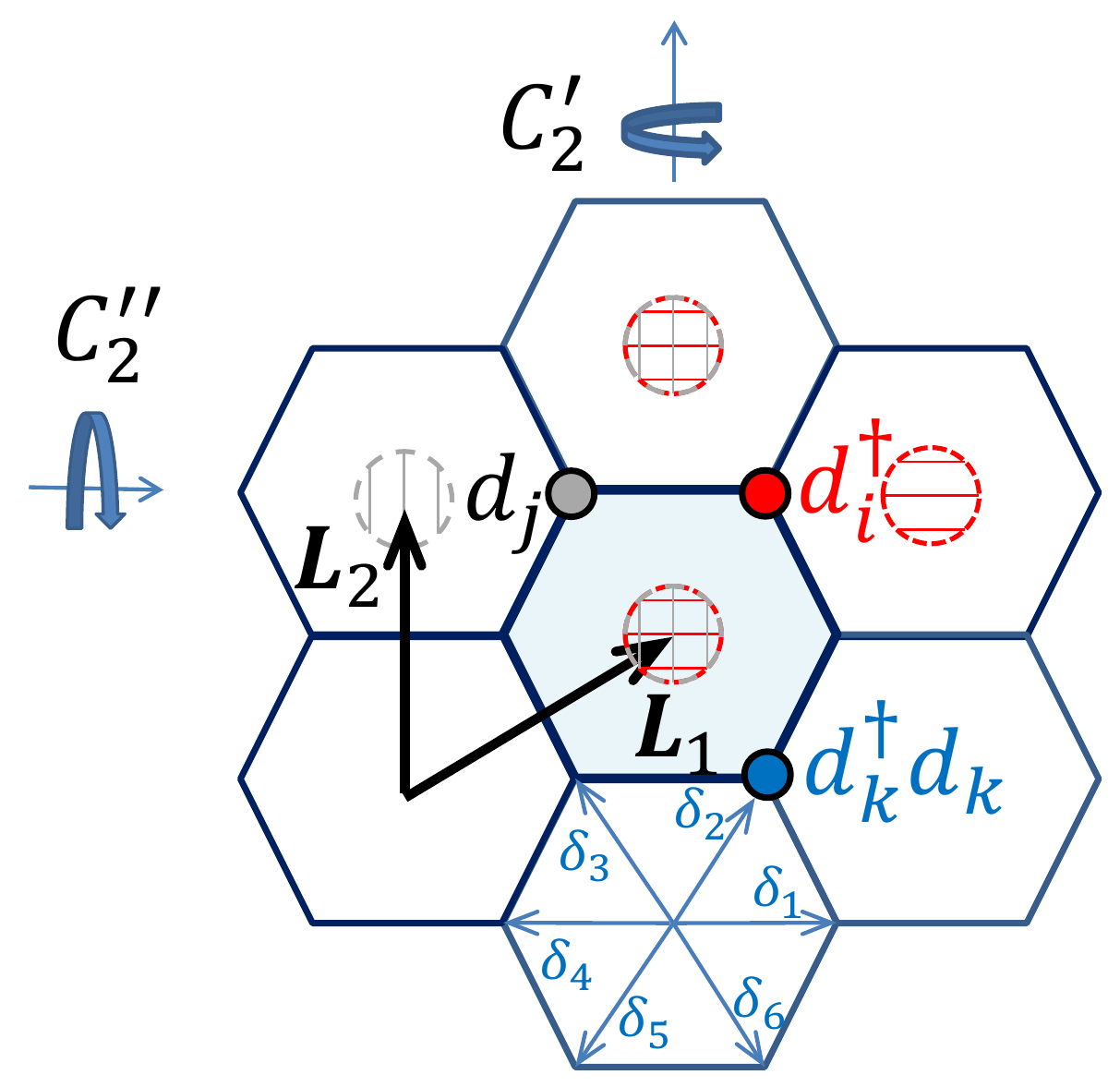}
\caption{The centers of the hexagons correspond to the triangular moire lattice spanned by primitive vectors ${\bf L}_{1,2}$. The Wannier state (WS) wavefunction centered on the moire honeycomb site $j$ has three peaks at the neighboring triangular moire sites (grey circles with vertical stripes). The WS on the neighboring site $i$ overlaps with it on the two hexagons (red horizontal stripes). An example of a four fermion interaction term, which is beyond the extended Hubbard model, appearing in the strong coupling Hamiltonian Eqs.(\ref{Eq:Uprojected},\ref{Eqn:OHam}-\ref{Eqn:THam}), is also shown schematically.}
\label{Fig:lattice}
\end{figure}

Here we present the analysis and the solution to the strong coupling limit by projecting the Coulomb interaction onto the microscopically constructed exponentially localized Wannier states (WSs) for the four narrow bands~\cite{KangVafek}. In doing so we find that there is a qualitative difference between the effect of the interactions in twisted bilayer graphene narrow bands and the much studied narrow band whose width is small due to the exponentially vanishing overlap of the well separated localized orbitals i.e. a solid in an atomic limit. In contrast, the small bandwidth in twisted bilayer graphene is a result of fine tuning (twist angle or pressure) and subtle interference of the WSs, and, unlike in the atomic limit, it is not necessarily a result of large spatial separation of the exponentially localized WSs. Indeed, as shown before, each WS of the twisted bilayer graphene narrow bands has three main peaks on neighboring sites of the triangular moire superlattice~\cite{AdrianPo,KangVafek,Koshino}. Therefore, for nearest neighbor WSs on say, sites $i$ and $j$, two peaks overlap significantly (see Fig.\ref{Fig:lattice}). Even though the integral under both has to vanish by orthogonality, the integral under each separately does not. This leads to a dramatically new form of the interaction Hamiltonian projected onto the narrow band basis  -- containing terms beyond the ``cluster Hubbard'' term~\cite{AdrianPo,Patrick} -- which in turn leads to different strong coupling phases as in the atomic limit. Specifically, the usual anti-ferromagnetic super-exchange mechanism fails and turns ferromagnetic. Due to approximate spin-valley $SU(4)$ symmetry, the fully spin-valley polarized ferromagnet is found to be degenerate with a spin-valley entangled state whose average total magnetic moment per particle vanishes. We also find exact excited states, which are spatially extended, and whose gap is suppressed by Zeeman coupling to an external magnetic field, making it (or at least its order parameter) a candidate for the experimentally observed correlated insulator at $2$ electrons/holes per moire unit cell.
At 1 particle per moire unit cell we find that the projected interactions favor an insulating stripe $SU(4)$ ferromagnet.
This state may be a candidate for the insulator observed at the $1/8$ filling\cite{CoryAndrea} if the $SU(4)$ degeneracy is lifted in favor of the physical spin ferromagnet.

We start by writing the full Hamiltonian as
\begin{eqnarray}\label{Eq:Hstart}
H &=& K+U,
\end{eqnarray}
where the kinetic energy $K$ is described by the tight-binding model~\cite{KangVafek} based on the WSs and where the Coulomb interaction is
\begin{eqnarray}
U &=& \frac{1}{2}\sum_{{\bf r},{\bf r'}}\sum_{\sigma,\sigma'=\uparrow,\downarrow}c^{\dagger}_\sigma({\bf r})c_\sigma({\bf r})V({\bf r-r'})c^{\dagger}_{\sigma'}({\bf r'})c_{\sigma'}({\bf r'}) \ .
\end{eqnarray}
Projecting onto the four narrow bands is equivalent to expanding $c_{\sigma}({\bf r})$ solely in terms of the narrow bands WSs
\begin{eqnarray}\label{Eq:cOp}
c_\sigma({\bf r})&=&\frac{1}{3}\sum_{{\bf R}}\sum_{p=1}^6\sum_{j=\pm1} w_{{\bf R}+\delta_p,j}({\bf r})d_{j,\sigma}({\bf R}+\delta_p) \ ,
\end{eqnarray}
where integers $m$, $n$ define the triangular moire lattice vectors ${\bf R}=m{\bf L}_1 + n{\bf L}_2$, the eigenvalue of the AA site centered 3-fold rotation $\exp(j2\pi i/3)$ is labeled by $j=\pm1$ and $\delta_{1,\ldots,6}$ are basis vectors connecting the honeycomb sites to the triangular sites (see Fig.\ref{Fig:lattice}). To an excellent approximation, WSs with $j=\pm1$ correspond to different valleys with very little valley mixing\cite{KangVafek}. The factor of $1/3$ is due to each honeycomb site position $\bR+ \fvec \delta_p$ being counted three times.

The Coulomb interaction $V(\br)$ is screened due to the presence of the metallic gates\cite{Pablo1,Pablo2,CoryAndrea}. The separation between the gates sets the length-scale beyond which the image charges exponentially diminish the repulsion\cite{Vafek15}. Interestingly, the gate separation is comparable to the moire unit cell. This, as well as the form of $w_{{\bf R}+ \fvec \delta_p,j}({\bf r})$ justifies keeping only $\bR=\bR'$ in the sum below:
\begin{widetext}
\begin{eqnarray}
U &=& \frac{1}{2}\sum_{\bR,\bR'}\sum_{{\bf r},{\bf r'}\in \hexagon}\sum_{\sigma,\sigma'=\uparrow,\downarrow}n_\sigma(\bR+\br)V(\bR+\br-\bR'-\br')
n_{\sigma'}(\bR'+\br')\\
&\approx&\frac{1}{2}\sum_{\bR}\sum_{{\bf r},{\bf r'}\in \hexagon}\sum_{\sigma,\sigma'=\uparrow,\downarrow}n_\sigma(\bR+\br)V(\br-\br')n_{\sigma'}(\bR'+\br'),
\end{eqnarray}
\end{widetext}
where $n_\sigma(\br)=c^{\dagger}_\sigma({\bf r})c_\sigma({\bf r})$ and the sums over $\br,\br'$ are restricted to be within the moire hexagon centered at the origin (see shaded $\hexagon$ in Fig.\ref{Fig:lattice}).

Substituting the Eq.(\ref{Eq:cOp}) into the above form, with numerically calculated $w_{{\bf R}+\delta_p,j}({\bf r})$ from the microscopic model~\cite{KangVafek} we find that to an excellent approximation we can replace $V(\br-\br')$ by its average over a region of size set by the extent of $w_{\delta_p,j}({\bf r})$ within the moire hexagon $V_0$, and because $V(\br)$ is dominated by the small wavevectors, we can ignore the valley mixing terms~\cite{Koshino}. Thus,
\begin{eqnarray}\label{Eq:Uprojected}
U&\approx&\frac{V_0}{2}\sum_{\bR}\left(\sum_{j=\pm1}\sum_{\sigma=\uparrow,\downarrow}O_{j,\sigma}(\bR)\right)^2,
\end{eqnarray}
where $O_{j,\sigma}(\bR)=\sum_{\br \in \hexagon}n_{j,\sigma}(\bR+\br)$ and
\begin{widetext}
\begin{eqnarray}
n_{j,\sigma}(\bR+\br)&=&\frac{1}{9}\sum_{\bar{\bR},\bar{\bR}'}\sum_{p,p'=1}^6
w^*_{\bar{\bR}-\bR+\delta_p,j}(\br)w_{\bar{\bR}'-\bR+\delta_{p'},j}(\br)d^\dagger_{j,\sigma}(\bar{\bR}+\delta_p)d_{j,\sigma}(\bar{\bR}'+\delta_{p'})\\
&\approx&\sum_{p,p'=1}^6
w^*_{\delta_p,j}(\br)w_{\delta_{p'},j}(\br)d^\dagger_{j,\sigma}(\bR+\delta_p)d_{j,\sigma}(\bR+\delta_{p'}) \ .
\end{eqnarray}
\end{widetext}
It is clear that $O_{j,\sigma}(\bR)$ is a superposition of not only density-like operators $d^\dagger_{j,\sigma}(\bR+\delta_p)d_{j,\sigma}(\bR+\delta_{p'})$ with $p=p'$, but also hopping-like terms with $p\neq p'$ which may be of the same order of magnitude. For example, $\sum_{\br\in\hexagon} w^*_{\delta_p,j}(\br)w_{\delta_{p+1},j}(\br)$ is non-negligible. This is despite the WSs being orthogonal when $\br$ is summed over {\it all} space; with $\br$ restricted to only one hexagon, the sum is $\mathcal{O}(1)$. For fixed $\bR$, the orthogonality in turn forces terms such as those with $p=1$ and $p'=2$ to be negative of the terms with $p=5$ and $p'=4$, etc. In what follows, we assume for clarity that the 3 peaks of each WS reside entirely within the 3 neighboring hexagons with no support elsewhere. We relax this assumption in the Supplementary material without any change to our conclusions~\cite{Supp}.
To summarize,
\begin{align}
  O_{j, \sigma}(\fvec R) & = \frac{1}{3} Q_{j, \sigma}(\fvec R) + \alpha_1 T_{j, \sigma}(\fvec R) \ , \quad \mbox{where}  \label{Eqn:OHam} \\
  Q_{j, \sigma}(\fvec R) & = \sum_{p = 1}^6 d^{\dagger}_{j, \sigma}(\fvec R + \fvec \delta_p) d_{j, \sigma}(\fvec R + \fvec \delta_p), \label{Eqn:QHam} \\
  T_{j, \sigma}(\fvec R ) & = \sum_{p = 1}^6 \left( e^{i \eta_{p,j}} d^{\dagger}_{j,\sigma}(\fvec R + \fvec \delta_{p+1}) d_{j,\sigma}(\fvec R + \fvec \delta_p)+h.c. \right), \label{Eqn:THam}
\end{align}
where $e^{i \eta_{p,j}}=(-)^{p-1} e^{i (-)^{p-1} \theta_j}$, $\fvec\delta_7=\fvec\delta_1$,
and
\begin{eqnarray}
  \alpha_1 e^{i \theta_j} & = \sum_{\fvec r \in \hexagon} w^*_{\fvec R + \fvec \delta_2, j, \sigma}(\fvec r) w_{\fvec R + \fvec \delta_1, j, \sigma}(\fvec r).
\end{eqnarray}
$\alpha_1 e^{i \theta_j}$ is generally a complex number and $\theta_{+1} = - \theta_{-1}$. This phase factor can be absorbed by applying a global $U(1)$ transformation on WSs. In the rest of the paper, we will therefore assume $\theta_{+1} = - \theta_{-1} = 0$. For our WSs constructed from the projection method~\cite{Supp}, $\alpha_1 \approx 0.23$.  Although not all the above interaction terms have been included in the model of Ref.~\cite{Koshino}, and although the Coulomb interaction is not assumed screened in Ref.\cite{Koshino}, similar value for $\alpha_1$ can be estimated from their ratio of the nearest-neighbor exchange and nearest neighbor density repulsion as $\alpha^{(K)}_1 \approx \frac{1}{3}\sqrt{J_1/V_1} \approx 0.16$ (see Table I of Ref.\cite{Koshino}). The nature of the ground state in the strong coupling limit is insensitive to such differences.

We emphasize that it is not necessary to include the kinetic energy terms $K$ in Eq.(\ref{Eq:Hstart}) to induce correlation among various sites; such sizable value of $\alpha_1$ makes the projected interaction term (\ref{Eq:Uprojected}) non-local \emph{even in the strong coupling limit}, and as we will see it dictates the  nature of the ground state.
It is therefore worth understanding why $\alpha_1$ is sizable. In the atomic limit, this overlap is exponentially small. As a consequence, the interactions usually include only the on-site terms, giving rise to the Hubbard model; $\alpha_1$ would then be set by the ratio of the bandwidth and the on-site repulsion. In our case, as mentioned, the two of the three peaks of the neighboring WSs spatially overlap and $\alpha_1\sim{\mathcal O}(1)$. This stems from the fact that the emergent two-fold symmetry $C_2''$ (see Fig.\ref{Fig:lattice}) is not locally implemented for our valley filtered WSs~\cite{SenthilTop}. Otherwise, when combined with (locally implemented) $C_2'$ (see Fig.\ref{Fig:lattice}) and the emergent valley $U(1)$ symmetry, all the WSs would have to have the same parity under $C_2''$\cite{SenthilTop}, leading to $\alpha_1 = 0$. However, $C_2''$ cannot be locally implemented simultaneously with the valley $U(1)$, $C_2'$, and the time reversal symmetry\cite{AdrianPo,SenthilTop}. $\alpha_1\sim\mathcal{O}(1)$ is thus rooted in the non-trivial topological properties of the narrow bands\cite{AdrianPo,SenthilTop,SenthilTop2,andrei,BJYang,Dai}.

\begin{figure}[htbp]
\centering
\subfigure[\label{Fig:oneQuarter}]{\includegraphics[width=0.45\columnwidth]{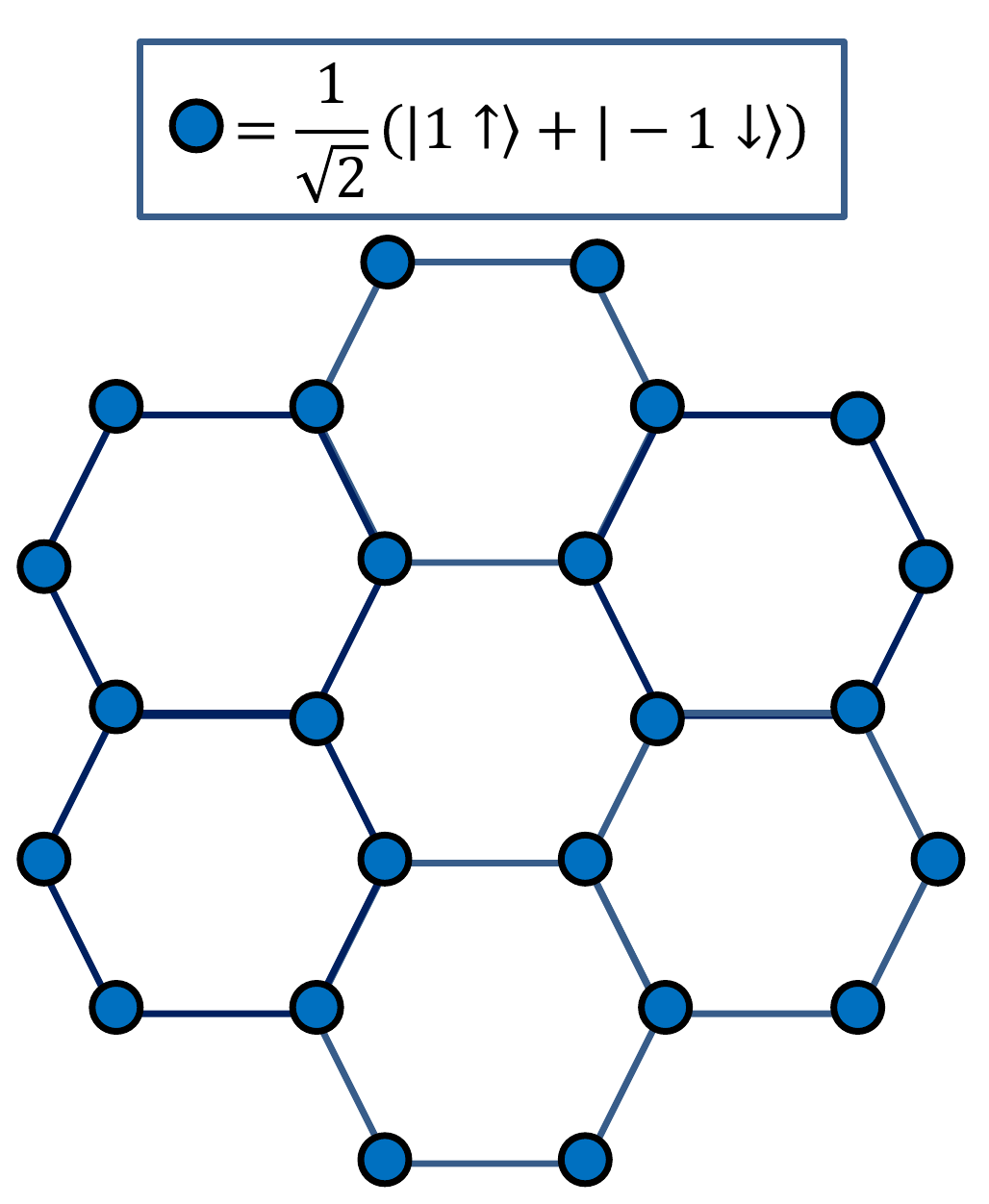}}
\subfigure[\label{Fig:oneEigth}]{\includegraphics[width=0.45\columnwidth]{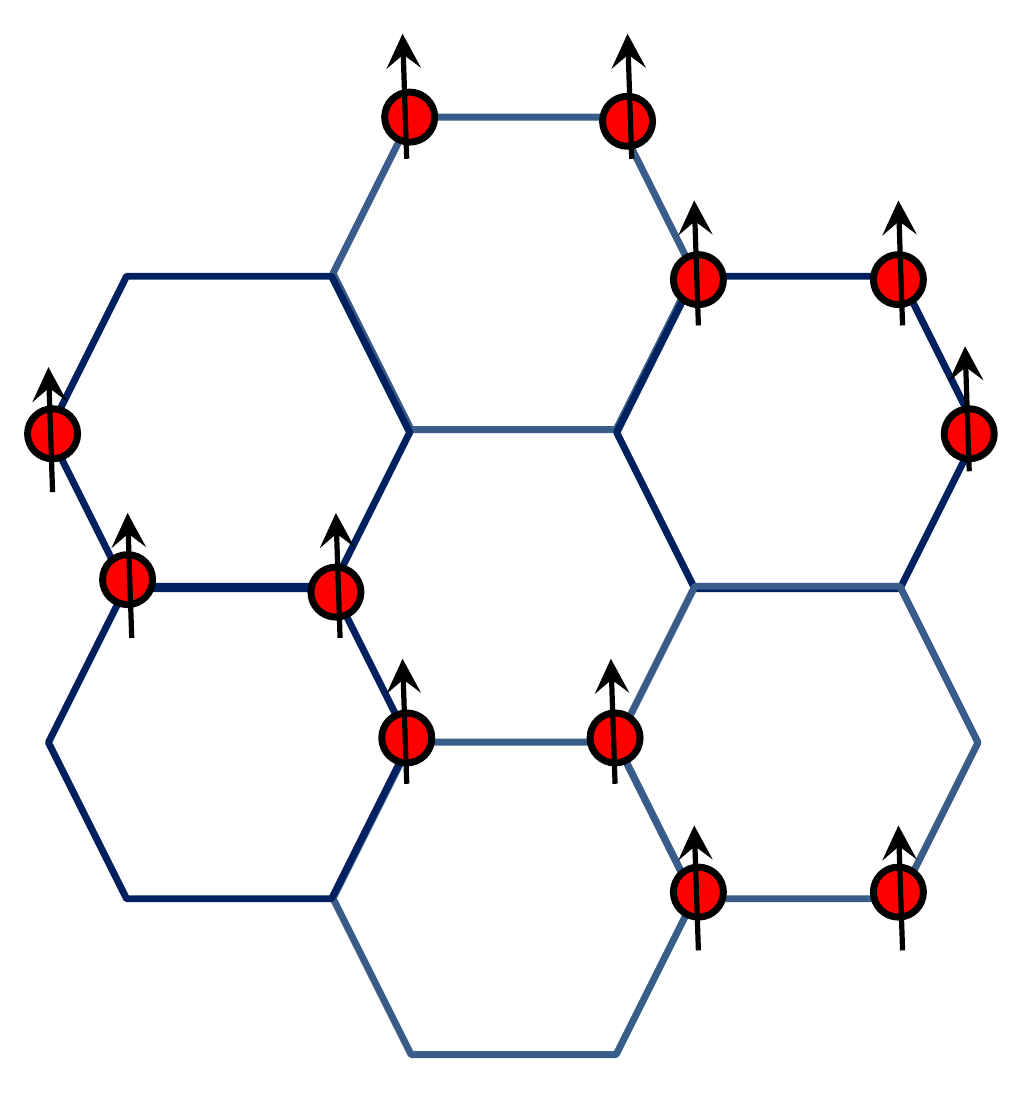}}
\caption{Schematic of the ground states at (a) $1/4$ filling ($2$ electrons/holes per moire unit cell) and at (b) $1/8$ filling ($1$ electron/hole per moire unit cell).}
\label{Fig:GS}
\end{figure}

As the first step, we therefore need to find the spectrum of the interaction $U$ in Eqn.~(\ref{Eq:Uprojected}).
This is non-trivial because the commutator $[O_{\sigma,j}(\bR),O_{\sigma,j}(\bR')]$ does not vanish for nearest neighbors $\bR$ and $\bR'$ due to $\alpha_1\neq 0$. However, the ground state of (\ref{Eq:Uprojected}) can be exactly solved for special fillings, including $2$ particles/holes per unit cell.
To see this, note that $\sum_\bR
\sum_j \sum_\sigma O_{j,\sigma}(\bR)=\hat{N}$, where $\hat{N}$ is the total particle number operator.
Therefore, we can write (\ref{Eq:Uprojected}) exactly as
\begin{eqnarray}
\frac{V_0}{2}\sum_{\bR}\left(n_0-\sum_{j, \sigma} O_{j,\sigma}(\bR)\right)^2+V_0n_0\hat{N}-\frac{V_0}{2}n^2_0N_\bR \label{Eqn:trick}
\end{eqnarray}
where $N_\bR$ is the total number of moire unit cells. Because $\hat{N}$ is fixed in the quantum number sector of interest, the last two terms are fixed.
The ground state thus minimizes the first term. But the first term is a sum of squares of Hermitian operators, and if we can find a state in which each term vanishes, we find the ground state. Let $n_0 = 2$. Then the state
\begin{eqnarray}
|\Phi_0\rangle = \prod_\bR d^{\dagger}_{j=1,\uparrow}(\bR + \fvec \delta_1)d^{\dagger}_{j=1,\uparrow}(\bR + \fvec \delta_2)|0\rangle
\end{eqnarray}
makes the first term vanish for every $\bR$, and is therefore a ground state.
This state corresponds to a fully spin/valley polarized ferromagnet with two electrons per moire unit cell.
Although it is a ground state, it is not the only one. Due to the $SU(4)$ symmetry of Eq.(\ref{Eq:Uprojected}), the ground state is $(2N_\bR+3)(2N_\bR+2)(2N_\bR+1)/6$ fold degenerate. This $SU(4)$ ground state manifold includes states as (see Fig.\ref{Fig:oneQuarter})
\begin{eqnarray}
|\Phi_1\rangle = \prod_\bR \prod_{p=1}^2 \frac1{\sqrt{2}} \left(d^{\dagger}_{1,\uparrow}(\bR+\delta_p)+d^{\dagger}_{-1,\downarrow}(\bR+\delta_p)\right)|0\rangle \ . \label{Eqn:GSNonmag}
\end{eqnarray}
Note that the expectation value of the square of the total (magnetic) spin  operator $\langle \Phi_1| \bS^2_{tot}|\Phi_1\rangle=\mathcal{O}(N_\bR)$, which means that the magnetic moment per particle vanishes in the thermodynamic limit. $|\Phi_1\rangle$ is therefore not a ferromagnet.

The ground state degeneracy is lifted by the kinetic terms, $K$ in Eq.(\ref{Eq:Hstart}), which in general break the $SU(4)$ symmetry. The valley $U(1)$ symmetric hopping terms
$t(\fvec R + \fvec \delta, \fvec R' + \fvec \delta')  d_{\fvec R + \fvec \delta, j, \sigma}^{\dagger} d_{\fvec R' + \fvec \delta', j, \sigma} $
favor the state with two valleys equally mixed, because then the second order process is least blocked. For the same reason the hopping terms that mix the valleys favor the state in which the two valleys carry opposite spins. The ground states is then given by Eqn.~\ref{Eqn:GSNonmag} up to a global spin $SU(2)$ rotation. The non-magnetic ground state depicted in Fig.\ref{Fig:oneQuarter} is thus favored by the kinetic terms.

We can also find some of the excited eigenstates of Eq.(\ref{Eq:Uprojected}) exactly.
In particular,
\begin{eqnarray}\label{Eq:excited1}
|N+1,j,\sigma; p\mod 2\rangle &=& \sum_{\bR} d^{\dagger}_{j,\sigma}(\bR+\delta_p) |\Phi_1\rangle,\\
|N-1,j,\sigma; p\mod 2\rangle &=& \sum_{\bR} d_{j,\sigma}(\bR+\delta_p) |\Phi_1\rangle,
\label{Eq:excited2}\end{eqnarray}
have energies $E_{N+1} = \frac{13}{6} V_0 +E_N$ and $E_{N-1} = - \frac{11}{6} V_0+E_N$, respectively, where $E_{N}=2 N_{\bR}V_0$. The gap is therefore at most $\Delta=E_{N+1}+E_{N-1}-2E_N = V_0/3$. Note that the excitations (\ref{Eq:excited1})-(\ref{Eq:excited2}) are spatially extended.

Even though the ground state $|\Phi_1\rangle$ does not couple linearly to the Zeeman magnetic field, ${\bf B}$, the excitations do, and the gap closes upon the application of a critical ${\bf B}$.

In order to gain some intuition for the physics behind the mathematical results discussed, imagine artificially tuning $\alpha_1$ to be small. At $\alpha_1=0$, ground states of the ``cluster Hubbard'' terms include states with one particle per honeycomb site. The small hopping terms give rise to exchange interactions $\mathcal{O}(\alpha_1^2)$, via both the usual second order perturbation theory and directly via the first order terms also of $\mathcal{O}(\alpha_1^2)$. The former would normally be anti-ferromagnetic, but in this case contributions from different hexagons cancel and only the latter, ferromagnetic exchange, remains.
The ground state manifold of the ``cluster Hubbard'' Hamiltonian also includes states which do not necessarily have one particle per site, but the same argument applies~\cite{Supp}.

Recent experiments also suggest that an insulating state appears at the filling of one hole/electron per unit cell, with the insulation enhanced by the Zeeman magnetic field\cite{CoryAndrea}. We were unable to find the exact ground state at this filling analytically, even in the strong coupling limit because $\alpha_1 \neq 0$. However, the ground state can be found if $\alpha_1$ is small. The leading term in $U$ is given by the ``cluster Hubbard'' terms, with ground states for which each hexagon contains three fermions, and $\sum_{j, \sigma} Q_{j, \sigma}(\fvec R) = 3$. Such ground states are highly degenerate even without counting the valley and spin degrees of freedom. The linear order and the second order of the cross term
$ \sum_{\fvec R} \left(\sum_{j, \sigma} Q_{j,\sigma}(\fvec R) \right) \left( \sum_{j', \sigma'} T_{j', \sigma'}(\fvec R) \right) $
vanish for the same reason as discussed above. Therefore, to the order $O(\alpha_1^2)$, only the term $\sum_{\fvec R} \left( \sum_{j, \sigma} T_{j,\sigma}(\fvec R) \right)^2$ contributes. This contribution is minimized if \emph{(1)} each hexagon contains exactly three occupied sites; \emph{(2)} each occupied site is in the same state; \emph{(3)} the number of bonds connecting an occupied site and an unoccupied site is minimized. These constraints favor the stripe $SU(4)$ ferromagnetic phase as the ground state, see Fig.~\ref{Fig:oneEigth}, with the energy correction $\delta E = \alpha_1^2 N_R V_0/2$. This phase is also an insulator due to the existence of the charge gap.

To summarize, we analysed the Coulomb interactions (screened by the gates) projected to the exponentially localized Wannier states\cite{KangVafek} for the four narrow bands in the ``magic angle'' twisted bilayer graphene. The projected interaction is highly non-local and is beyond extended Hubbard models. Such novel interactions result from the non-trivial topological properties of the narrow bands\cite{AdrianPo,SenthilTop}, giving rise to the $SU(4)$ ferromagnetic ground states at $1/4$ and $1/8$ filings. At $1/4$ filling, the kinetic terms break the $SU(4)$ symmetry and select the state in which two valleys with opposite spins are equally mixed (Fig.~\ref{Fig:oneQuarter}). This state, although still $SU(4)$ ferromagnetic, is (physical) spin non-magnetic in the thermodynamic limit, with a charge gap suppressed by the magnetic field. We also argue that the stripe $SU(4)$ ferromagnetic insulator phase is the ground state at $1/8$ filling (Fig.~\ref{Fig:oneEigth}). If the $SU(4)$ degeneracy is lifted in favor of the physical spin ferromagnet, such state could be a candidate for the experimentally observed insulator at the $1/8$ filling~\cite{CoryAndrea}. The mechanism of such symmetry breaking remains an open problem.

\begin{acknowledgments}
JK was supported by the National High Magnetic Field Laboratory through NSF Grant No.~DMR-1157490 and the State of Florida.  O. V. was supported by NSF DMR-1506756.
\end{acknowledgments}

\newpage
\widetext \vspace{0.5cm}

\begin{center}
\textbf{\large{}{}{}Supplementary Material for ``Strong coupling phases of partially filled twisted bilayer graphene narrow bands{}''}{\large{}{} }
\par\end{center}

\setcounter{equation}{0} \setcounter{figure}{0} \setcounter{subfigure}{0}
\setcounter{table}{0} \makeatletter \global\long\def\theequation{S\arabic{equation}}
 \global\long\def\thefigure{S\arabic{figure}}  \global\long\def\thetable{S\arabic{table}}
\makeatother

\global\long\def\bibnumfmt#1{[S#1]}
 \global\long\def\citenumfont#1{S#1}

\section{Construction of the Wannier States}
\begin{figure}[htbp]
\centering
\subfigure[\label{FigS:TBG}]{\includegraphics[width=0.35\columnwidth]{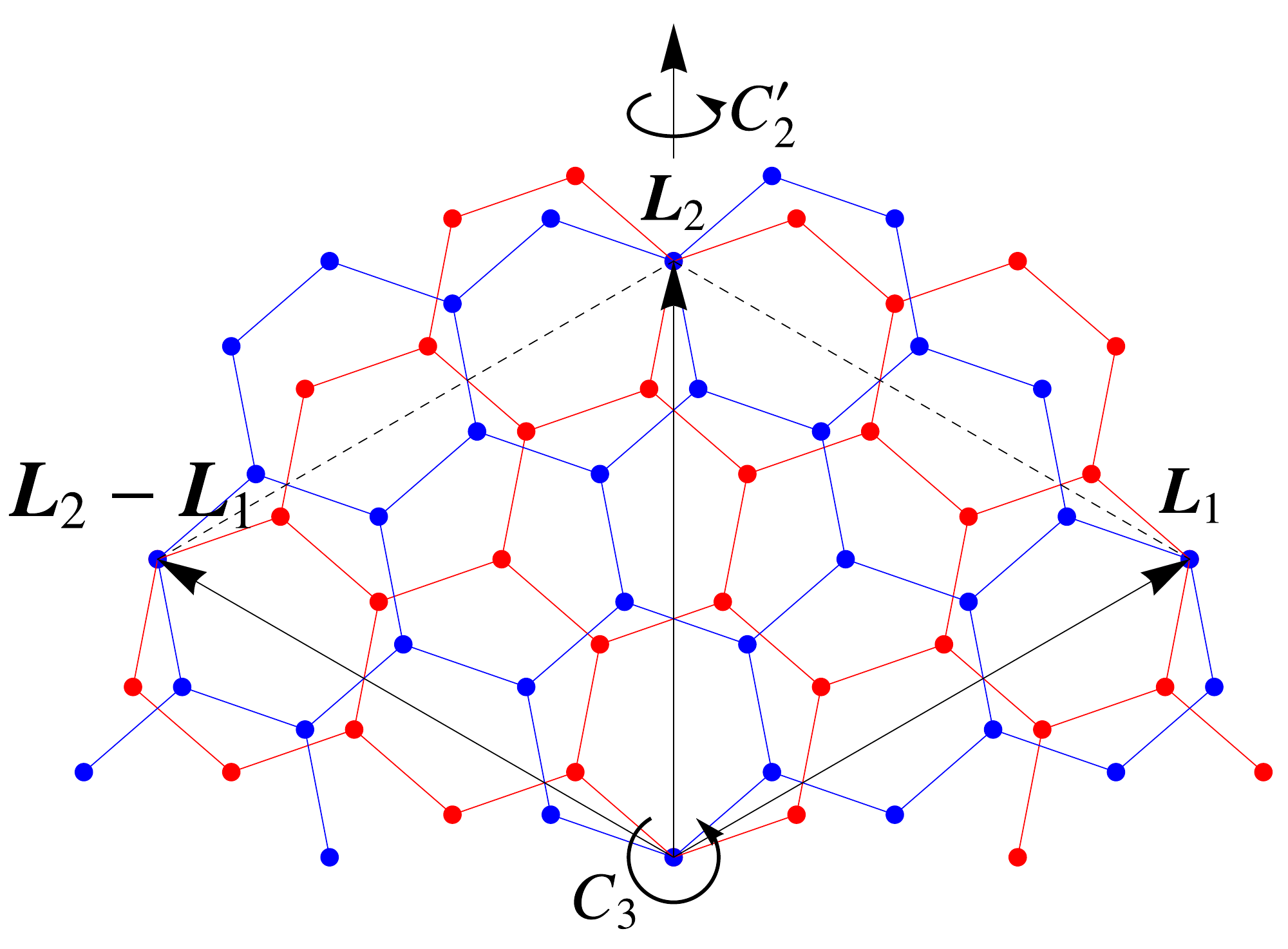}}
\subfigure[\label{FigS:SValDensity}]{\includegraphics[width=0.3\columnwidth]{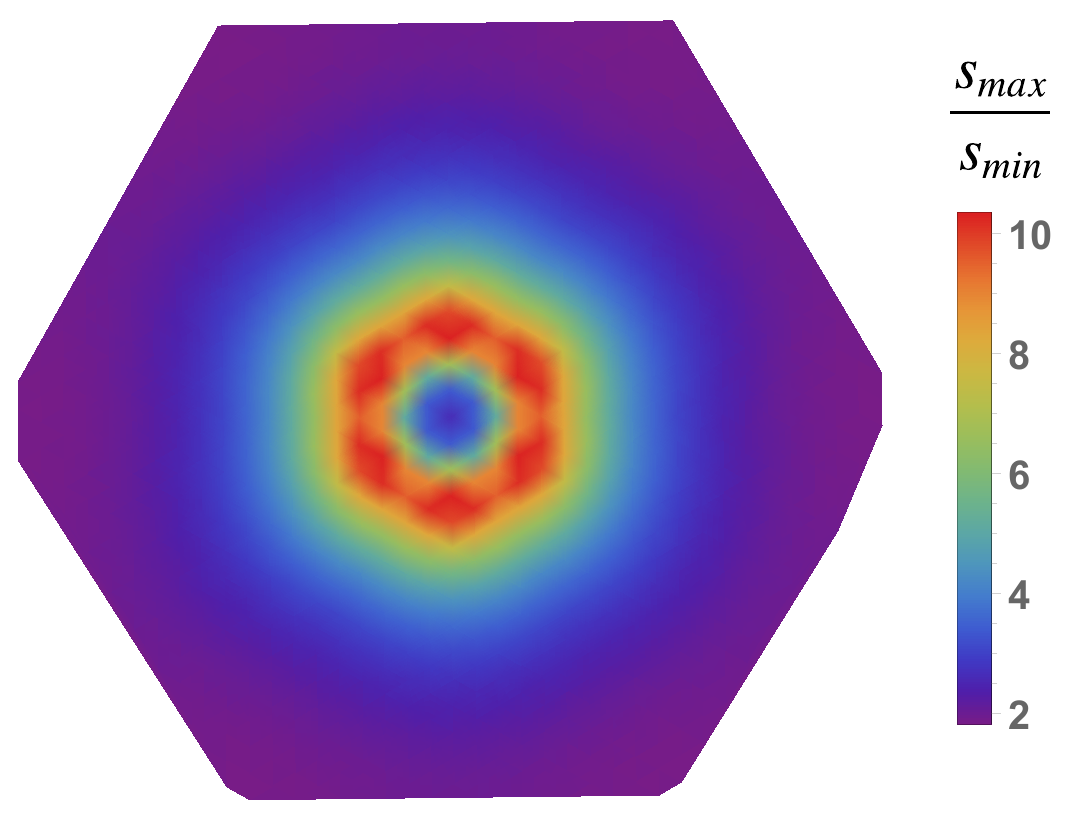}}
\subfigure[\label{FigS:SValHeight}]{\includegraphics[width=0.32\columnwidth]{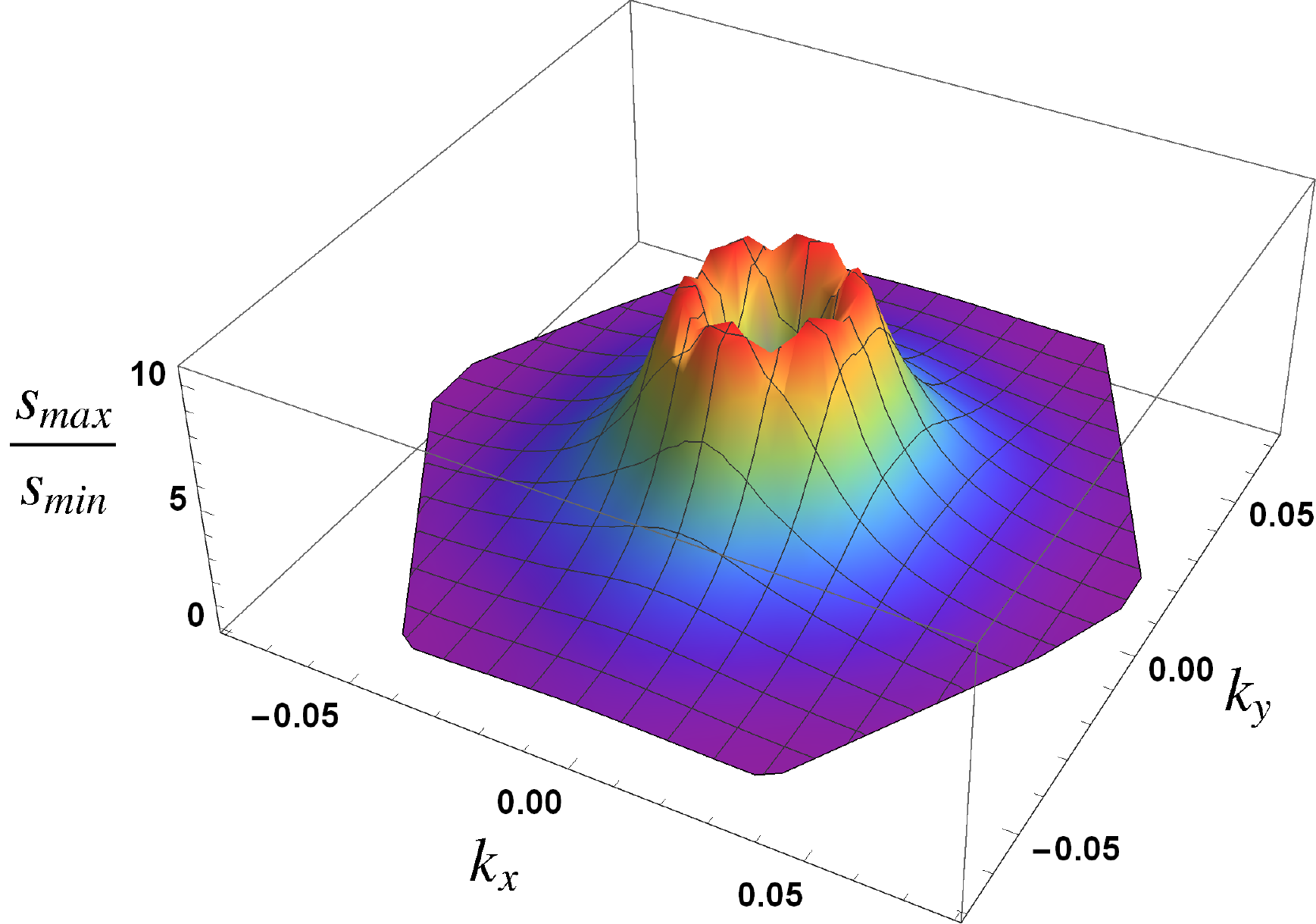}}
\caption{(a) The schematic plot of the twisted bilayer graphene. (b) The ratio between the maximal singular value and the minimal one in the momentum space. (c) the same plot as (b) but shows the ratio in $\hat z$ direction. It is clear that this ratio is always below $10$ but greater than $2$.}
\label{FigS:SValRatio}
\end{figure}

In our study, we construct the Wannier states (WSs) of the twisted bilayer graphene with the twist angle of $\sim1.3^{\circ}$ and $m = 25$ $n =26$ (for details and notation, see Ref.\cite{KangVafekS}). We will follow the projection method~\cite{WSProjS} and choose the initial ansatz to have the same symmetry as the final WSs. Although very similar to what has been done in~\cite{KangVafekS}, the initial ansatz here is chosen slightly differently in order to improve the localization of WSs and to maintain the nearly perfect valley polarization:
\begin{itemize}
  \item $h_1$: As shown in Fig.~\ref{FigS:TBG}, our $h_1$ is defined as  $\Psi_{\Gamma, E_+, \epsilon}$ only on sublattice A inside the triangle $\fvec 0 - \fvec L_1 - \fvec L_2$, and $0$ otherwise. This choice guarantees that $h_1$ transform in the same way  as $\Psi_{\Gamma, E_+, \epsilon}$ (with $C_3$ the eigenvalue of $\epsilon = \exp(i 2\pi/3)$) under the three-fold rotation around the center of the triangle.
  \item $h_2$: Apply complex conjugation to $h_1$. Therefore, $h_1$ and $h_2$ transform to each other under time reversal, and $h_2$ has the eigenvalue of $\epsilon^*$ under the three-fold  rotation around the center of the triangle.
  \item $h_3$: Apply $C_2'$ to $h_1$. Note that $h_3$ is nonzero only inside the triangle $\fvec 0 - (\fvec L_2- \fvec L_1) - \fvec L_2$. In addition, $h_3$ has the eigenvalue of $\epsilon^*$  under the three-fold rotation around the center of the triangle.
  \item $h_4$: Apply $C_2'$ to $h_2$. It is obvious that $h_4$ and $h_3$ transform to each other under time reversal. In addition, $h_4$ has the eigenvalue of $\epsilon$ under the three-fold rotation around the center of the triangle $\fvec 0 - (\fvec L_2- \fvec L_1) - \fvec L_2$.
\end{itemize}
This ansatz is chosen to improve the localization of the WSs obtained from the projection method.

Fig.~\ref{FigS:SValRatio} shows the ratio between the maximal and minimal singular values of the matrix $A(\fvec k)_{ij} = \langle \Psi_i(\fvec k) | h_j \rangle$ as a function of momentum~\cite{WSProjS,KangVafekS}, where $\Psi_i(\fvec k)$ is the Bloch state at the momentum of $\fvec k$. This ratio would become infinite if the matrix $A(\fvec k)$ were singular, and that would lead to delocalized WSs; if this cannot be avoided for any choice of $h_{1}$,$h_{2}$,$h_{3}$ and $h_{4}$, then there is an obstruction\cite{WSProjS}. As shown in Fig.~\ref{FigS:SValRatio}, the matrix is never singular for our choice of the initial ansatz.

It can be shown that the symmetry of the WSs after the projection method is the same as that of the initial ansatz states $h_j$.

\section{Screened Coulomb Potential}
In this section, we will present our numerical result to confirm that the projected Coulomb potential can be written in the form of Eqn.~\ref{Eq:Uprojected}
with the constraint that $\sum_{\fvec R, j,\sigma} \hat{O}_{j,\sigma}(\fvec R) = \hat{N}$.

As explained in the text, the Coulomb potential is screened due to the image charges induced by two gates above and below the twisted bilayer graphene. As derived in Ref.~\cite{Vafek15S}, the screened Coulomb interaction is given by
\begin{equation}
  V(\fvec r) = U_{\xi}  \sum_{n = - \infty}^{\infty} \frac{(-1)^n}{\sqrt{ (r/\xi)^2 + n^2 }} \ , \label{EqnS:Coulomb}
\end{equation}
where $\xi \approx 10$nm~\cite{Pablo1S,Pablo2S,CoryAndreaS} is half of the distance between two gates, and $\epsilon \approx 6$ is the dielectric constant of BN. This leads to $U_{\xi} = e^2/(4\pi \epsilon \xi) = 24$meV. Eqn.~\ref{EqnS:Coulomb} works as long as $\fvec r \neq 0$.  In our formula, the on-site repulsion ($\fvec r =0$) is set to be $2 \times 9.3 = 18.6$eV~\cite{Blugel11S}. Since the Coulomb potential decays exponentially for $|\fvec r| > \xi$~\cite{Vafek15S} and projected $n(\fvec r)$ is concentrated around the center of the hexagon, we consider only the interactions that $\fvec r$ and $\fvec r'$ are located in the same hexagon. Therefore, the Coulomb interaction is well approximated by
\begin{align}
  U & = \frac12 \sum_{\fvec R} \sum_{\sigma \sigma'} \sum_{\fvec r \fvec r' \in \hexagon}  V(\fvec r - \fvec r') n_{\sigma}(\fvec R + \fvec r) n_{\sigma'}(\fvec R + \fvec r').
\end{align}
We next project the fermion number operator $n(\fvec r)$ to the WSs for the narrow bands. As each site contains four different states with $j = \pm 1$ and $\sigma = \uparrow \downarrow$,
\[ n_{\sigma}(\fvec r \in \hexagon) \rightarrow \sum_{j j'} \sum_{\fvec R, \fvec \delta} \sum_{\fvec R', \fvec \delta'} w^*_{\fvec R + \fvec \delta, j}(\fvec r) w_{\fvec R' + \fvec \delta', j'}(\fvec r)  d^{\dagger}_{j, \sigma}(\fvec R + \fvec \delta) d_{j', \sigma}(\fvec R' + \fvec \delta') \ , \]
where $\fvec R$ ($\fvec R'$) specifies the unit cell and $\fvec \delta$ ($\fvec \delta'$) $= \fvec \delta_1$ or $\fvec \delta_2$ refers to the two honeycomb lattice sites in the unit cell (illustrated in Fig.~\ref{Fig:lattice}). If each WS has only three peaks well localized around the neighboring triangular lattice sites, the projected $n(\fvec r)$ is dominated by the WS at the sites of the hexagon. In this supplementary material, however, we will not make this assumption and argue that our conclusions in the main text still holds.

\subsection{Cluster Hubbard terms}
As explained in the main text, the cluster Hubbard terms can be written as
\begin{align}
  U_0 & = \half \sum_{\sigma, \sigma'} \sum_{j, j'} \sum_{\fvec R} \sum_{p,p' = 1}^6 V^{(0)}_{p, j; p', j'} \rho_{j,\sigma}(\fvec R + \fvec \delta_p) \rho_{j',\sigma'}(\fvec R + \fvec \delta_{p'}) \quad \mbox{where} \\
  V^{(0)}_{p, j; p', j'} & = \sum_{\fvec r, \fvec r' \in \hexagon}  V(\fvec r - \fvec r') |w_{\fvec \delta_p, j}(\fvec r)|^2 |w_{\fvec \delta_{p'}, j'}(\fvec r')|^2 \ .
\end{align}
$\rho_{j,\sigma}(\fvec R + \fvec \delta_p) = d^{\dagger}_{j, \sigma}(\fvec R + \fvec \delta_p) d_{j, \sigma}(\fvec R + \fvec \delta_p)$ is the projected on-site fermion number operator. Since $w_{\fvec \delta_p, 1}(\fvec r) = \big( w_{\fvec \delta_p, -1}(\fvec r) \big)^*$, the interaction constant $V^{(0)}_{p, j; p', j'}$ is independent of the valley indices $j$ and $j'$. In the following, we simplify the notation $V^{(0)}_{j,p; j', p'} = V^{(0)}_{p p'}$. The on-site charging interaction constants are given in Tab.~\ref{tabS:OnSite}:
\begin{table}[h]
\begin{tabular}{|c|c|c|c|c|}
\hline
 $V_{11}^{(0)}$ & $V_{12}^{(0)}$ & $V_{13}^{(0)}$ & $V_{14}^{(0)}$ & $V_{23}^{(0)}$   \\ \hline
 $9.03$ & $8.92$ & $8.79$ & $8.75$ & $8.90$   \\ \hline
\end{tabular}
\caption{The on-site charging interactions. All other on-site interaction constants can be obtained by the symmetry transformation $C_3$ and $C_2'$. All numbers here are in the unit of meV.}
\label{tabS:OnSite}
\end{table}

All these numbers are almost identical. Thus, we set $V^{(0)}_{pp'} \approx V^{(0)} = 8.87$meV.

\subsection{Assisted Nearest Neighbor Hopping}
As explained in the main text, the interaction term also includes the assisted nearest neighbor hopping terms. The additional interaction terms can be grouped into two parts: the cross terms between the nearest neighbor hopping and the on-site density, and the square of the nearest hopping:
\begin{align}
  U_1 & = \frac12 \sum_{\fvec R} \sum_{p,p'} \sum_{j, j'} \sum_{\sigma, \sigma'} \left\{ \left( V^{(1)}_{p, j; p', j'} \rho_{j, \sigma}(\fvec R + \fvec \delta_p) d^{\dagger}_{j', \sigma'}(\fvec R + \fvec \delta_{p'})  d_{j', \sigma'}(\fvec R + \fvec \delta_{p'+1}) + h.c.  \right) +  \right. \nonumber \\
   &  \left( V^{(2)}_{p, j; p', j'} d^{\dagger}_{j, \sigma}(\fvec R + \delta_{p})  d_{j, \sigma}(\fvec R + \fvec \delta_{p+1})  d^{\dagger}_{j', \sigma'}(\fvec R + \fvec \delta_{p'})  d_{j', \sigma'}(\fvec R + \fvec \delta_{p'+1}) + h.c. \right) + \nonumber \\
   &  \left. \left( V^{(3)}_{p, j; p' j'} d^{\dagger}_{j, \sigma}(\fvec R + \fvec \delta_{p})  d_{j, \sigma}(\fvec R + \fvec \delta_{p+1})  d^{\dagger}_{j', \sigma'}(\fvec R + \fvec \delta_{p'+1})  d_{j', \sigma'}(\fvec R + \fvec \delta_{p'}) + h.c. \right)  \right\}   \label{EqnS:NNInt}
\end{align}
with the definition of $\fvec \delta_7 \equiv \fvec \delta_1$. Same convention will be used in our paper for notation convenience. These interaction constants are calculated as
\begin{align}
  V^{(1)}_{p, j; p', j'} & = \sum_{\fvec r, \fvec r' \in \hexagon}   V(\fvec r - \fvec r') \left| w_{\fvec \delta_p, j}(\fvec r) \right|^2 w_{\fvec \delta_{p'}, j'}^*(\fvec r') w_{\fvec \delta_{p'+1}, j'}(\fvec r') \\
  V^{(2)}_{p, j; p', j'} & = \sum_{\fvec r, \fvec r' \in \hexagon}  V(\fvec r - \fvec r') w^*_{\fvec \delta_p, j}(\fvec r)  w_{\fvec \delta_{p+1}, j}(\fvec r)  w^*_{\fvec \delta_{p'}, j'}(\fvec r')  w_{\fvec \delta_{p' + 1}, j'}(\fvec r') \\
  V^{(3)}_{p, j; p', j'} & = \sum_{\fvec r, \fvec r' \in \hexagon}  V(\fvec r - \fvec r') w^*_{\fvec \delta_{p+1}, j}(\fvec r)  w_{\fvec \delta_p, j}(\fvec r)   w^*_{\fvec \delta_{p'}, j'}(\fvec r')  w_{\fvec \delta_{p' + 1}, j'}(\fvec r').
\end{align}
It is obvious that the first interaction constants $V^{(1)}$ is independent of the indices $j$ and $V^{(1)}_{p, j; p', j'} = ( V^{(1)}_{p, j; p', -j'} )^*$. So we will drop the index $j$ in $V^{(1)}$. These interaction constants are listed in Table.~\ref{tabS:NN1}.

\begin{table}[h]
\begin{tabular}{|c|c|c|c|c|c|}
\hline
 $V^{(1)}_{1;1\epsilon}$ & $V^{(1)}_{1;2\epsilon}$ & $V^{(1)}_{1;3\epsilon}$ & $V^{(1)}_{1;4\epsilon}$ & $V^{(1)}_{1;5\epsilon}$ & $V^{(1)}_{1;6\epsilon}$  \\ \hline
 $7.49 e^{0.744\pi i}$ & $7.54 e^{0.249\pi i}$ & $7.29 e^{0.748\pi i}$ & $7.46 e^{0.249\pi i}$ & $7.36 e^{0.748\pi i}$ & $7.63 e^{0.247\pi i}$  \\ \hline
\end{tabular}
\caption{ The crossing term between the on-site density and the nearest neighbor hopping. Other interaction constants of the crossing term can be obtained by the symmetry transformation $C_3$, $C_2'$, and time reversal. All numbers here are in the unit of meV. }
\label{tabS:NN1}
\end{table}
Therefore, these cross terms can be  approximated as
\begin{align}
    V^{(1)} \sum_{\fvec R} & \left( \sum_{j', \sigma'} \sum_{p' = 1,3,5}  e^{i \theta j'} d^{\dagger}_{j', \sigma'}(\fvec R + \fvec \delta_p') d_{j', \sigma'}(\fvec R + \fvec \delta_{p' + 1})  \right. \nonumber \\
    & \left. \ - e^{-i\theta j'} d^{\dagger}_{j', \sigma'}(\fvec R + \fvec \delta_{p' + 1}) d_{j', \sigma'}(\fvec R + \fvec \delta_{p' + 2}) +  h.c. \right) \left( \sum_{p,\sigma, j} \rho_{j, \sigma}(\fvec R + \fvec \delta_p) \right)
\end{align}
with $\theta \approx 0.75\pi$ and $V^{(1)} \approx 7.46$meV.

\begin{table}[h]
\begin{tabular}{|c|c|c|c|c|c|}
\hline
 $V^{(2)}_{1\epsilon;1\epsilon}$ & $V^{(2)}_{1\epsilon;2\epsilon}$ & $V^{(2)}_{1\epsilon;3\epsilon}$ & $V^{(2)}_{1\epsilon;4\epsilon}$ & $V^{(2)}_{2\epsilon;2\epsilon}$ & $V^{(2)}_{2\epsilon;4\epsilon}$  \\ \hline
 $6.14 e^{-0.510\pi i}$ & $6.26 e^{0.995\pi i}$ & $6.10 e^{-0.505\pi i}$ & $6.22 e^{0.995\pi i}$ & $6.40 e^{0.495\pi i}$ & $6.39 e^{0.497\pi i}$   \\ \hline
 $V^{(2)}_{1\epsilon;1\epsilon^*}$ & $V^{(2)}_{1\epsilon;2\epsilon^*}$ & $V^{(2)}_{1\epsilon;3\epsilon^*}$ &  $V^{(2)}_{1\epsilon;4\epsilon^*}$ & $V^{(2)}_{2\epsilon;2\epsilon^*}$ & $V^{(2)}_{2\epsilon;4\epsilon^*}$ \\ \hline
 $6.41$ & $6.39 e^{0.498\pi \ii}$ & $6.08$ & $6.18 e^{0.499\pi i}$ & $6.48$ & $6.37$  \\ \hline\hline
  $V^{(3)}_{1\epsilon;1\epsilon}$ & $V^{(3)}_{1\epsilon;2\epsilon}$ & $V^{(3)}_{1\epsilon;3\epsilon}$ & $V^{(3)}_{1\epsilon;4\epsilon}$ & $V^{(3)}_{2\epsilon;2\epsilon}$ & $V^{(3)}_{2\epsilon;4\epsilon}$  \\ \hline
 $6.39$ & $6.38 e^{0.498\pi i}$ & $6.07$ & $6.16 e^{0.499\pi i}$ & $6.46$ & $6.36$   \\ \hline
  $V^{(3)}_{1\epsilon;1\epsilon^*}$ & $V^{(3)}_{1\epsilon;2\epsilon^*}$ & $V^{(3)}_{1\epsilon;3\epsilon^*}$ &  $V^{(3)}_{1\epsilon;4\epsilon^*}$ & $V^{(3)}_{2\epsilon;2\epsilon^*}$ & $V^{(3)}_{2\epsilon;4\epsilon^*}$ \\ \hline
 $6.14 e^{-0.510\pi i}$ & $6.27 e^{0.995\pi \ii}$ & $6.10 e^{-0.505\pi i}$ & $6.22 e^{0.995\pi i}$ & $6.40 e^{0.495\pi i}$ & $6.39 e^{0.497\pi i}$  \\ \hline
\end{tabular}
\caption{ The square of the nearest neighbor hopping. Other  interaction constants can be obtained by the symmetry transformation $C_3$, $C_2'$, and time reversal. All numbers here are in the unit of meV. }
\label{tabS:NN2}
\end{table}

Again, the interaction constants in Tab.~\ref{tabS:NN2} suggest that the 2nd and 3rd terms in Eqn.~\ref{EqnS:NNInt} can be well approximated as
\[  \frac{V^{(2)}}2 \sum_{\fvec R} \left( \sum_{j', \sigma'} \sum_{p' = 1,3,5} \left( e^{i \theta j'} d^{\dagger}_{j', \sigma'}(\fvec R + \fvec \delta_p') d_{j', \sigma'}(\fvec R + \fvec \delta_{p' + 1})  - e^{-i\theta j'} d^{\dagger}_{j', \sigma'}(\fvec R + \fvec \delta_{p' + 1}) d_{j', \sigma'}(\fvec R + \fvec \delta_{p' + 2}) \right) +  h.c. \right)^2    \]
with $V^{(2)} \approx 6.28$meV. It is interesting that $V_2  V_0 / (V_1)^2 = 1.001 \approx 1$. This suggests that the whole interaction can be written in a simple form:
\begin{align}
  U & = \frac{V_0}2 \sum_{\fvec R} \left( \alpha_0 Q(\fvec R) + \alpha_1 T(\fvec R) \right)^2   \quad \mbox{where} \\
  Q(\fvec R) & = \sum_{p, j, \sigma} \rho_{j,\sigma}(\fvec R + \fvec \delta_p) \label{EqnS:Onsite} \\
  T_1(\fvec R) & = \sum_{p = 1, 3, 5} \sum_{j, \sigma} \left(  e^{i \theta j} d^{\dagger}_{j, \sigma}(\fvec R + \fvec \delta_p) d_{j, \sigma}(\fvec R + \fvec \delta_{p + 1})  - e^{-i\theta j} d^{\dagger}_{j, \sigma}(\fvec R + \fvec \delta_{p + 1}) d_{j, \sigma}(\fvec R + \fvec \delta_{p + 2}) +  h.c. \right)
\end{align}
In addition, we note that the ratio between two coefficients $\alpha_1 / \alpha_0 = V_1/V_0 = 0.84$, very close to the ratio of two WS overlaps inside the hexagon:
\[  t_0 = \sum_{\fvec r \in \hexagon} |w_1(\fvec r)|^2 = 0.286 \ , \quad t_1 = \sum_{\fvec r \in \hexagon} w_1^*(\fvec r) w_4(\fvec r) = 0.23 e^{0.743\pi i}  \ , \quad t_1/t_0 = 0.81 e^{0.743\pi i} \ .  \]
We see that the phase $\theta \approx Arg[t_1]$ and $|t_1|/t_0 \approx \alpha_1 / \alpha_0$, suggesting that the approximation introduced in the main text agrees with our numerical calculation very well. Therefore, it is natural to choose that
\begin{align}
  \alpha_0 & = t_0 \ , \alpha_1 = |t_1|  \quad \Longrightarrow  \quad V_0 = V^{(0)}/\alpha_0^2 = 108\mathrm{meV} \ , \qquad  U  = \frac{V_0}2 \sum_{\fvec R} \big( \alpha_0 Q(\fvec R) + \alpha_1 T_1(\fvec R) \big)^2.
\end{align}
Note that the phase $\theta \neq 0$, but a gauge transformation can be applied to absorb it:
\[ \left\{ \begin{array}{lll}
  d_{j, \sigma}(\fvec R + \fvec \delta_p) \longrightarrow e^{i \theta j/2}  d_{j, \sigma}(\fvec R + \fvec \delta_p) \quad & \mbox{for } p = 1,\ 3,\ 5   \\
  d_{j, \sigma}(\fvec R + \fvec \delta_p) \longrightarrow e^{-i \theta j/2} d_{j, \sigma}(\fvec R + \fvec \delta_p)  \quad & \mbox{for } p = 2,\ 4,\ 6 \   .
\end{array}  \right.  \]
After applying this gauge transformation, we see that
\begin{align}
    T_1(\fvec R) & = \sum_{p = 1, 3, 5} \sum_{j, \sigma}  \left( d_{j, \sigma}^{\dagger}(\fvec R + \fvec \delta_p) d_{j, \sigma}(\fvec R + \fvec \delta_{p+1}) - d_{j, \sigma}^{\dagger}(\fvec R + \fvec \delta_{p+1}) d_{j, \sigma}(\fvec R + \fvec \delta_{p+2})  \right)  \ .  \label{EqnS:NNHop}
\end{align}

\subsection{Assisted Next-nearest Neighbor Hopping}
As explained in the text, the assisted hopping between next-nearest neighbor is small because of the orthogonality between two overlapped peaks. However, our numerical calculation shows that this term is still significant because our constructed WSs also contains several secondary peaks, giving rise to the sizable overlap between the next nearest neighbor WSs within the hexagon:
\[  \alpha_2 = \sum_{\fvec r \in \hexagon} w^*_{1, \fvec \delta_p}(\fvec r) w_{1,\fvec \delta_{p+2}}(\fvec r) \approx  -0.13 \ .  \]
The interaction, when including this next-nearest neighbor assisted hopping, can still be written in a similar form:
\begin{align}
  U & = \frac{V_0}2 \sum_{\fvec R} \big( \alpha_0 Q(\fvec R) + \alpha_1 T_1(\fvec R) + \alpha_2 T_2(\fvec R) \big)^2  \quad \mbox{where} \quad
   T_2(\fvec R) = \sum_{p, \sigma} \sum_{j = \pm 1}  d^{\dagger}_{j, \sigma}(\fvec R + \fvec \delta_p) d_{j, \sigma}(\fvec R + \fvec \delta_{p+2}) + h.c.
\end{align}
We found $\sum_{\fvec R} \alpha_0 Q(\fvec R) + \alpha_1 T_1(\fvec R) + \alpha_2 T_2(\fvec R) \not\approx \hat{N}$. This relation can be recovered by including the next nearest hopping beyond the hexagon. Here, we define
\begin{align}
  T_2'(\fvec R) & = \sum_{j, \sigma} d_{j, \sigma}^{\dagger}(\fvec R + \fvec \delta_1) \big( d_{j,\sigma}(\fvec R +\fvec \delta_1 + \fvec L_1) + d_{j,\sigma}(\fvec R +\fvec \delta_1 + \fvec L_1 - \fvec L_2)  \big) + \nonumber \\
  & \hspace{1.0cm}  d_{j, \sigma}^{\dagger}(\fvec R + \fvec \delta_2) \big( d_{j,\sigma}(\fvec R +\fvec \delta_2 + \fvec L_2) + d_{j,\sigma}(\fvec R +\fvec \delta_2 + \fvec L_1)  \big) + \nonumber \\
  & \hspace{1cm} d_{j, \sigma}^{\dagger}(\fvec R + \fvec \delta_3) \big( d_{j,\sigma}(\fvec R +\fvec \delta_3 + \fvec L_2) + d_{j,\sigma}(\fvec R +\fvec \delta_3 + \fvec L_2 - \fvec L_1)  \big) + \nonumber \\
  & \hspace{1cm} d_{j, \sigma}^{\dagger}(\fvec R + \fvec \delta_4) \big( d_{j,\sigma}(\fvec R +\fvec \delta_4 - \fvec L_1) + d_{j,\sigma}(\fvec R + \fvec \delta_4 -\fvec L_1 + \fvec L_2)  \big) + \nonumber \\
  & \hspace{1cm} d_{j, \sigma}^{\dagger}(\fvec R + \fvec \delta_5) \big( d_{j,\sigma}(\fvec R +\fvec \delta_5 - \fvec L_1) + d_{j,\sigma}(\fvec R + \fvec \delta_5 - \fvec L_2)  \big) + \nonumber \\
  & \hspace{1cm} d_{j, \sigma}^{\dagger}(\fvec R + \fvec \delta_6) \big( d_{j,\sigma}(\fvec R +\fvec \delta_6 - \fvec L_2) + d_{j,\sigma}(\fvec R + \fvec \delta_6 - \fvec L_2 + \fvec L_1 )  \big) + h.c.
\end{align}
The interaction also includes the contribution from this additional next nearest neighbor hopping term:
\begin{align}
  U & = V_0 \sum_{\fvec R} \big( \alpha_0 Q(\fvec R) +\alpha_1 T_1(\fvec R) + \alpha_2 T_2(\fvec R) + \alpha_2' T'_2(\fvec R)  \big)^2  \\
  \alpha_2' & \approx \sum_{\fvec r\in \hexagon} w_{\fvec \delta_1}^*(\fvec r)  w_{\fvec L_1 + \fvec \delta_1}(\fvec r) \approx 0.05
\end{align}
Now, when summing over all the hexagons, we have
\[  \sum_{\fvec R} \alpha_0 Q(\fvec R) + \alpha_1 T_1(\fvec R) + \alpha_2 T_2(\fvec R) + \alpha_2' T_2'(\fvec R) \approx 3 \alpha_0 \hat{N} \approx \hat{N}  \]
This relation becomes almost exact when including more hopping terms beyond the hexagon. For simplicity, we stop here and assume that the interaction is
\[  U  = \frac{V_0}2 \sum_{\fvec R} \left( \alpha_0 Q(\fvec R) + \alpha_1 T_1(\fvec R) + \alpha_2 T_2(\fvec R) + \alpha_2' T_2'(\fvec R)  \right)^2     \]
with $\alpha_0 = 1/3$ and $\alpha_2' = - \alpha_2/2$, so that
\[   \sum_{\fvec R} \alpha_0 Q(\fvec R) + \alpha_1 T_1(\fvec R) + \alpha_2 T_2(\fvec R) + \alpha_2' T_2'(\fvec R) = \hat{N} \ .  \]

\section{Ground State and Excited States at $1/4$ Filling}
As explained in the main text, we will follow the strong coupling approach, ie.~to treat the hopping term as perturbation and find the ground states at $1/4$ filling. It turns out that the product state
\[ \big| \Phi_{GS} \rangle = \prod_{\fvec R, \fvec \delta} d^{\dagger}_{j, \sigma_{\fvec n}}(\fvec R + \fvec \delta) | 0 \rangle  \]
is the ground state of the interaction $U$. $\fvec \delta$ refers to the two hexagon sites in each unit cell, and the creation operator $d^{\dagger}_{j, \sigma_{\fvec n}}$ creates a fermion with the valley $j$ and spin along the direction of $\fvec n$. Since the interaction $U$ is $SU(4)$ symmetric, any hoppings that conserve both spin and valley annihilate  $| \Phi_{GS} \rangle$ because the state on each honeycomb lattice site is identical. Therefore, for any hexagon,
\[  \sum_{j, \sigma}O_{j, \sigma}(\fvec R) \big| \Phi_{GS} \rangle = 6 \alpha_0  \quad \Longrightarrow \quad E_{GS} = 18 N_R \alpha_0^2 V_0 \ , \]
where $N_R$ is the number of the unit cell. It seems that the energy of the ground state depends on the parameter $\alpha_0$, which depends on the constructed WSs. However, when including more cluster Hubbard terms from WSs located outside the hexagon,
\[ \alpha_0 + \alpha_0' + \cdots = \frac13  \quad \Longrightarrow \quad   E_{GS} = 2 N_R V_0 \ .      \]


\subsection{$SU(4)$ Symmetry Breaking}
In this subsection, we will discuss how the $SU(4)$ degeneracy can be lifted by the kinetic terms in the Hamiltonian. The valley $U(1)$ symmetric hopping terms can be generally written as~\cite{KangVafekS}
\begin{align}
  & K(\fvec R + \fvec \delta, \fvec R' + \fvec \delta')  = \sum_{\sigma} \left( t(\fvec R + \fvec \delta, \fvec R' + \fvec \delta') d^{\dagger}_{1, \sigma}(\fvec R + \fvec \delta)  d_{1, \sigma}(\fvec R' + \fvec \delta') + \right. \nonumber \\
   & \qquad \left. t^*(\fvec R + \fvec \delta, \fvec R' + \fvec \delta') d^{\dagger}_{-1, \sigma}(\fvec R + \fvec \delta)  d_{-1, \sigma}(\fvec R' + \fvec \delta')  \right) + h.c. \nonumber \\
  & K = \sum_{\fvec R, \fvec \delta} \sum_{\fvec R', \fvec \delta'} K(\fvec R + \fvec \delta, \fvec R' + \fvec \delta') \label{EqnS:Hopping}
\end{align}

The most general form of the ground state  is
\begin{align}
 | \Phi_{GS} \rangle = \prod_{\fvec R, \fvec \delta} \frac1{\sqrt{1+|\beta|^2}} \left( d^{\dagger}_{1, \sigma_{\fvec n_1}}(\fvec R + \fvec \delta) + \beta\  d^{\dagger}_{-1, \sigma_{\fvec n_2}}(\fvec R + \fvec \delta)  \right) | 0 \rangle \ , \label{EqnS:GSform}
\end{align}
where $\fvec n_1$ ($\fvec n_2$) are the direction of the spin polarizations. Note that this is the most general form for the ground state of the interaction $U$. When $\beta = 0$ or $\infty$, the state is both valley and spin polarized. If $|\beta| = 1$, the state mixes both valleys with equal weights.

In out approach, the hopping term in Eqn.~\ref{EqnS:Hopping} is treated as perturbation. It is clear that the first order perturbation vanishes as $\langle \Phi_{GS} | K | \Phi_{GS} \rangle = 0$ since any hopping changed the fermion number in two or more hexagons. The second order perturbation gives the correction as
\begin{align}
  \delta E^{(2)}_{GS} = - \sum_{\Phi_{ex}} \frac{|\langle \Phi_{ex} | K | \Phi_{GS} \rangle|^2}{|E_{GS} - E_{ex}|} \ , \label{EqnS:EHop2}
\end{align}
where $\sum_{\Phi_{ex}}$ sums over all the excited states. Since the spectrum of the excited states is almost impossible to solve, we instead maximize
\begin{align}
  & \sum_{\Phi_{ex}} |\langle \Phi_{ex} | K | \Phi_{GS} \rangle|^2 =  \left\Vert \sum_{\fvec R + \fvec \delta, \fvec R' + \fvec \delta'} K(\fvec R + \fvec \delta, \fvec R' + \fvec \delta') | \Phi_{GS} \rangle \right\Vert^2 = \sum_{\fvec R + \fvec \delta, \fvec R' + \fvec \delta'} \left\Vert K(\fvec R + \fvec \delta, \fvec R' + \fvec \delta') | \Phi_{GS} \rangle \right\Vert^2 \nonumber \\
  = & \sum_{\fvec R + \fvec \delta, \fvec R' + \fvec \delta'} \frac{|\beta|^2}{(1 + |\beta|^2)^2} 4 \big( Im(t(\fvec R + \fvec \delta, \fvec R' + \fvec \delta') ) \big)^2 \leq \sum_{\fvec R + \fvec \delta, \fvec R' + \fvec \delta'} \big( Im(t(\fvec R + \fvec \delta, \fvec R' + \fvec \delta') ) \big)^2
\end{align}
where $\Vert\cdots \Vert$ is the norm of the state. Since the hopping constants $t$ are in general complex numbers, $|\beta| = 1$ to maximize the norm. Thus, the ground state is given by the mixture of two valleys with equal weights.

Furthermore, in the presence of the valley mixing hopping terms, the kinetic terms contain
\[  K_{il}' = t_{il}' d^{\dagger}_{1, \sigma}(\fvec R_i + \fvec \delta_i) d_{-1, \sigma}(\fvec R_l + \fvec \delta_l) + ( t_{il}')^* d^{\dagger}_{-1, \sigma}(\fvec R_i + \fvec \delta_i) d_{1, \sigma}(\fvec R_l + \fvec \delta_l)   \]
following the same approach, we maximize the norm of
\[  \left\Vert K' | \Phi_{GS} \rangle \right\Vert^2 = \sum_{i l} \frac{|t_{il}'|^2}{(1 + |\beta|^2)^2} \left[ (1 + |\beta|^2)^2 - 2|\beta|^2 |\langle \sigma_{n_1}| \sigma_{n_2} \rangle |^2 \right]   \]
This norm is maximized  if and only if  $\langle \sigma_{\fvec n_1} | \sigma_{\fvec n_2} \rangle = 0$, ie.~$\fvec n_1 = - \fvec n_2 = \fvec n$. Thus, the ground state is
\[  \prod_{\fvec R, \fvec \delta} \frac1{\sqrt{2}} \left( d^{\dagger}_{1, \sigma_{\fvec n}}(\fvec R + \fvec \delta) + e^{i\theta} d^{\dagger}_{-1, \sigma_{- \fvec n}}(\fvec R + \fvec \delta)   \right)| 0 \rangle  \]
Note that the ground state is $SU(2)$ degenerate in spin space since the spin orientation $\fvec n$ is not fixed and the phase $\theta$ is arbitrary. For simplicity, we assume $\fvec n = \hat{z}$ and set $\theta = 0$.
 \[  | \Phi_{GS} \rangle = \prod_{\fvec R, \fvec \delta} \frac1{\sqrt{2}} \left( d^{\dagger}_{1, \uparrow}(\fvec R + \fvec \delta) + d^{\dagger}_{-1, \downarrow}(\fvec R + \fvec \delta)   \right) | 0 \rangle  \]
It is interesting to calculate the expectation value of $\fvec S^2$ where $\fvec S$ is the total spin:
\begin{align}
   \fvec S & = \sum_{\fvec r} c^{\dagger}_{\alpha}(\fvec r) \fvec \sigma_{\alpha\beta} c_{\beta}(\fvec r) \longrightarrow  \sum_{\fvec R} \sum_{p = 1}^2 \sum_j d_{j, \alpha}^{\dagger}(\fvec R + \fvec \delta_p)  \fvec \sigma_{\alpha\beta} d_{j, \beta}(\fvec R + \fvec \delta_p)
\end{align}
After projection to the WSs, the expectation value of $\fvec S^2$ for the ground state is:
\begin{align}
  \langle S_z^2 \rangle & = \sum_{\fvec R_i, \fvec \delta_i} \sum_{\fvec R_j, \fvec \delta_j} \langle \Phi_{GS} | S_z(\fvec R_i + \fvec \delta_i) S_z(\fvec R_j + \fvec \delta_j) | \Phi_{GS} \rangle
\end{align}
The formula above is nonzero only when $\fvec R_i + \fvec \delta_i = \fvec R_j + \fvec \delta_j$:
\[ \langle S_z^2 \rangle = \sum_{\fvec R, \fvec \delta} \langle \Phi_{GS} | S_z^2(\fvec R + \fvec \delta) | \Phi_{GS} \rangle = \frac{N_R}4 \ ,  \]
where $N$ is the number of honeycomb lattice sites. Similar results are obtained for $\langle S_x^2 \rangle$ and $\langle S_y^2 \rangle$. Therefore,
\[  \langle \fvec S^2 \rangle = \frac34 N_R \propto N_R  \quad \Longrightarrow \quad \frac{\sqrt{ \langle \fvec S^2 \rangle}}{2N_R} \propto N_R^{-\frac12}    \]
The average magnetic moment per particle is proportional to $1/\sqrt{N_R}$, thus vanishes in the thermodynamic limit.

\subsection{Excited States}
Consider the state
\begin{align}
  | \Phi_{N+1} \rangle = \frac1{\sqrt{N_R}} \sum_{\fvec R} d^{\dagger}_{1, \downarrow}(\fvec R + \fvec \delta_1) | \Phi_{GS} \rangle \   .  \label{EqnS:ExcNP1}
\end{align}
To show it is also the eigenstate of the interaction $U$, note that
\[  O(\fvec R) = \alpha_0 Q(\fvec R) + \alpha_1 T_1(\fvec R) + \alpha_2 T_2(\fvec R) + \alpha_2' T_2'(\fvec R)  \ .    \]
This leads to
\begin{align}
  & \big[ O(\fvec R),\ \frac1{\sqrt{N_R}} \sum_{\fvec R'} d^{\dagger}_{1, \downarrow}(\fvec R' + \fvec \delta_1) \big] \nonumber \\
  = & \frac1{\sqrt{N_R}} \left[ \sum_{p = 1, 3, 5} (\alpha_0 + 2\alpha_2 + 2 \alpha_2') d^{\dagger}_{1, \downarrow}(\fvec R + \fvec \delta_p) + \ \alpha_2' \left( d^{\dagger}_{1, \downarrow}(\fvec R +\fvec \delta_1 + \fvec L_1) + d^{\dagger}_{1,\downarrow}(\fvec R + \fvec \delta_1 + \fvec L_1 - \fvec L_2) +  \right. \right. \nonumber \\
  & \quad  \left. \left. d^{\dagger}_{1, \downarrow}(\fvec R +\fvec \delta_3 + \fvec L_2) + d^{\dagger}_{1,\downarrow}(\fvec R +\fvec \delta_3 + \fvec L_2 - \fvec L_1)  + d^{\dagger}_{1,\downarrow}(\fvec R + \fvec \delta_5 - \fvec L_1)  + d^{\dagger}_{1,\downarrow}(\fvec R + \fvec \delta_5 - \fvec L_2)  \right)  \right] \nonumber \\
  \Longrightarrow & \sum_{\fvec R} \big[ O(\fvec R),\ \frac1{\sqrt{N_R}}\sum_{\fvec R'} d^{\dagger}_{1,\downarrow}(\fvec R' + \fvec \delta_1) \big] =  \frac3{\sqrt{N_R}} \left( \alpha_0 + 2\alpha_2 + 4 \alpha_2' \right) \sum_{\fvec R'} d^{\dagger}_{1,\downarrow}(\fvec R' + \fvec \delta_1) \\
  & \Big[ O(\fvec R),\  \big[ O(\fvec R),\ \frac1{\sqrt{N_R}}\sum_{\fvec R'} d^{\dagger}_{1, \downarrow}(\fvec R' + \fvec \delta_1) \big] \Big] \nonumber \\
  = & \frac1{\sqrt{N_R}} \left[ \sum_{p = 1, 3, 5} \big[ (\alpha_0 + 2\alpha_2)(\alpha_0 + 2\alpha_2 + 2 \alpha_2') + 2\alpha_2'^2  \big] d^{\dagger}_{1, \downarrow}(\fvec R + \fvec \delta_p) +  \alpha_2' (\alpha_0 + 2\alpha_2 + 2 \alpha_2') \left( d^{\dagger}_{1, \downarrow}(\fvec R +\fvec \delta_1 + \fvec L_1) + \ \right. \right. \nonumber \\
  & \quad d^{\dagger}_{1,\downarrow}(\fvec R + \fvec \delta_1 + \fvec L_1 - \fvec L_2) + d^{\dagger}_{1, \downarrow}(\fvec R + \fvec \delta_3 + \fvec L_2) + d^{\dagger}_{1,\downarrow}(\fvec R +\fvec \delta_3 + \fvec L_2 - \fvec L_1)  + d^{\dagger}_{1, \downarrow}(\fvec R + \fvec \delta_5 - \fvec L_1)  + \nonumber \\
  & \left. \left. \quad d^{\dagger}_{1,\downarrow}(\fvec R + \fvec \delta_5 - \fvec L_2)  \right)  \right] \nonumber \\
  \Longrightarrow & \sum_{\fvec R} \Big[ O(\fvec R),\  \big[ O(\fvec R),\ \frac1{\sqrt{N_R}}\sum_{\fvec R'} d^{\dagger}_{1, \downarrow}(\fvec R' + \fvec \delta_1) \big] \Big] = \frac3{\sqrt{N_R}} \Big( \left( \alpha_0 + 2\alpha_2 + 2\alpha_2' \right)^2 + 2\alpha_2'^2 \Big) \sum_{\fvec R'} d^{\dagger}_{1, \downarrow}(\fvec R' + \fvec \delta_1)
\end{align}
The energy of the excited state is given as
\begin{align}
  & \left( \sum_{\fvec R} O(\fvec R)^2 \right) \frac1{\sqrt{N_R}} \sum_{\fvec R'} d_{1, \downarrow}^{\dagger}(\fvec R' + \fvec \delta_1) |\Phi_{GS} \rangle \nonumber \\
  = & \sum_{\fvec R}  \left( 2\left[ O(\fvec R), \   \frac1{\sqrt{N_R}} \sum_{\fvec R'} d_{1, \downarrow}^{\dagger}(\fvec R' + \fvec \delta_1)  \right] O(\fvec R) + \left[ O(\fvec R),\  \left[ O(\fvec R),\ \frac1{\sqrt{N_R}} \sum_{\fvec R'} d^{\dagger}_{1, \downarrow}(\fvec R' + \fvec \delta_1) \right] \right] + \right. \nonumber \\
   & \qquad  \left. \frac1{\sqrt{N_R}} \sum_{\fvec R'} d_{1, \downarrow}^{\dagger}(\fvec R' + \fvec \delta_1) O(\fvec R)^2 \right) | \Phi_{GS} \rangle \nonumber \\
  = & 6 (\alpha_0 + 2\alpha_2 + 4 \alpha_2') 6 \alpha_0 | \Phi_{N+1} \rangle + 3 \Big( \left( \alpha_0 + 2\alpha_2 + 2\alpha_2' \right)^2 + 2\alpha_2'^2 \Big) | \Phi_{N+1} \rangle   +  36 \alpha_0^2 N_R | \Phi_{N+1} \rangle \nonumber \\
  = & \left( 36\alpha_0 (\alpha_0 + 2\alpha_2 + 4 \alpha_2') + 3 \left( \alpha_0 + 2\alpha_2 + 2\alpha_2' \right)^2 + 6\alpha_2'^2  +  36 \alpha_0^2 N_R \right) | \Phi_{N+1} \rangle
\end{align}
Thus, we conclude that
\[  E_{N+1} = \frac{V_0}2 \left( 36 N_R \alpha_0^2 + 36\alpha_0 (\alpha_0 + 2\alpha_2 + 4 \alpha_2') + 3 \left( \alpha_0 + 2\alpha_2 + 2\alpha_2' \right)^2 + 6\alpha_2'^2 \right) \]
Note that the creation operator in Eqn.~\ref{EqnS:ExcNP1} is applied only on one sublattice. We can apply the same operator on another sublattice and obtain the eigenstate with the same energy $E_{N+1}$. In addition, the excited state could also be generated by the creation operator $d^{\dagger}_{-1, \uparrow}$ and $\frac1{\sqrt{2}}(d^{\dagger}_{1, \uparrow} - d^{\dagger}_{-1, \downarrow})$. Thus, there exist six extended states with the same energy.

For $| \Phi_{N-1} \rangle$ state, we consider
\[ | \Phi_{N-1} \rangle =  \frac1{\sqrt{N_R}} \sum_{\fvec R} \frac1{\sqrt{2}} \left( d_{1, \uparrow}(\fvec R + \fvec \delta_1) + d_{-1, \downarrow}(\fvec R + \fvec \delta_1) \right) | \Phi_{GS} \rangle \   .    \]
Following the same method, we conclude that this is also the eigenstate of the interaction $U$ with the energy of
\[  E_{N - 1} =  \frac{V_0}2 \left( 36 N \alpha_0^2 - 36\alpha_0 (\alpha_0 + 2\alpha_2 + 4 \alpha_2') + 3 \left( \alpha_0 + 2\alpha_2 + 2\alpha_2' \right)^2 + 6\alpha_2'^2 \right)   \]
Thus, the gap to the extended state is
\[ \Delta \leq E_{N+1} + E_{N-1} -2E_N = 3 V_0 \left( \left( \alpha_0 + 2\alpha_2 + 2\alpha_2' \right)^2 + 2 \alpha_2'^2 \right)   \]
For simplicity, we set $\alpha_2' = - \alpha_2/2$ and $\alpha_0 = 1/3$ to satisfy the constraint that $\sum_{\fvec R} O(\fvec R) = \hat{N}$. This leads to an upper limit of the gap:
\[ \Delta \leq 3 V_0 \left( \big(\frac13 + \alpha_2 \big)^2 + \frac{\alpha_2^2}2  \right)  \ . \]
Since $- 1/3 < \alpha_2 < 0$, this assisted next-nearest neighbor hopping term generally decrease the upper limit of the gap.

\section{Rise of SU($4$) Ferromagnetic Exchange Interaction}
In this section, we will discuss how the ferromagnetic interactions arise from the assistant hopping terms in $U$. For simplicity, we include only the on-site particle number $Q(\fvec R)$ and the nearest neighbor hopping $T_1$ and set $\alpha_0 = 1/3$. Thus
\[ U = \sum_{\fvec R} \left( O(\fvec R) \right)^2 \quad \mbox{where} \quad O(\fvec R) = \frac13 Q(\fvec R) + \alpha_1 T_1(\fvec R) \ .  \]
with $Q(\fvec R)$ and $T_1(\fvec R)$ defined in Eqn.~\ref{EqnS:Onsite} and \ref{EqnS:NNHop}. To understand how the ferromagnetic exchange interaction rises, consider the limit that $\alpha_1 \ll \alpha_0$ and treat the nearest neighbor assisted hopping terms as perturbation.
\begin{align}
  U & = U_0 + U_1 + U_2   \\
  U_0 & = \alpha_0^2 V_0 \sum_{\fvec R} \big( Q(\fvec R) \big)^2 \ , \\
  U_1 & = 2 \alpha_0 V_0 \alpha_1 \sum_{\fvec R} Q(\fvec R) T(\fvec R) \\
  U_2 & = \alpha_1^2 V_0 \sum_{\fvec R} \big( T(\fvec R) \big)^2 \ .
\end{align}
As explained in the main text, instead of considering the spectrum of $\sum_{\fvec R} \big( \alpha_0 Q(\fvec R) \big)^2$, we can study the ground states of $\sum_R \big( \alpha_0 Q(\fvec R) - 2\big)^2$ for $1/4$ filling. The lowest energy level is given by the state in which each hexagon contains $6$ fermions ($Q_{\fvec R} | \Phi_{GS} \rangle = 6 | \Phi_{GS} \rangle$). Such states are highly degenerate even without including the valley and spin degrees of freedom. To lift this degeneracy, we consider the pertubative expansion of the small coefficient $\alpha_1$. Note the cross term $Q(\fvec R) T(\fvec R)$ can be written as
\[  U_1  = 2 \alpha_0 \alpha_1 V_0 \sum_{\langle i j \rangle} \left( Q_{\fvec R_1} - Q_{\fvec R_2} \right) T_{ij} \ ,  \]
where $\langle i j \rangle$ is a nearest neighbor bond, and $\fvec R_1$ and $\fvec R_2$ refers to two hexagons that share this bond. $T_{ij}$ is the nearest hopping term on this bond. Therefore, the cross term $U_1$ annihilates all the states in the manifold of the ground states of $U_0$. As a consequence, the perturbative expansion of $U_1$ vanished up to the second order $\mathcal{O}(\alpha_1^2)$. The contribution of $U_2$ contains the term
\begin{align*}
   & \langle \Phi_{GS} | d^{\dagger}_{j,\sigma}(\fvec R + \fvec \delta_p) d_{j,\sigma}(\fvec R + \fvec \delta_{p + q}) d^{\dagger}_{j',\sigma'}(\fvec R + \fvec \delta_{p'}) d_{j',\sigma'}(\fvec R + \fvec \delta_{p' + q'}) | \Phi_{GS} \rangle
\end{align*}
where $q, q' = \pm 1$ for nearest neighbor hopping. It can be shown that this term is nonzero only when $p = p' + q'$ and $p' = p + q$. Therefore, we can focus only on the term
\[   d^{\dagger}_{j,\sigma}(\fvec R + \fvec \delta_p) d_{j,\sigma}(\fvec R + \fvec \delta_{p + q}) d^{\dagger}_{j',\sigma'}(\fvec R + \fvec \delta_{p+q}) d_{j',\sigma'}(\fvec R + \fvec \delta_{p})  \ .  \]
This term gives the $SU(4)$ ferromagnetic exchange. To be more explicit, this term can be written as the ferromagnetic coupling of two spin operators when $j = j'$ as follows:
\begin{align}
  & d^{\dagger}_{j,\sigma}(\fvec R + \fvec \delta_p) d_{j,\sigma}(\fvec R + \fvec \delta_{p + q}) d^{\dagger}_{j,\sigma'}(\fvec R + \fvec \delta_{p+q}) d_{j,\sigma'}(\fvec R + \fvec \delta_{p}) \nonumber \\
   = & -2 \fvec S_{j}(\fvec R + \fvec \delta_p) \cdot \fvec S_{j}(\fvec R + \fvec \delta_{p+q}) + n_j(\fvec R + \fvec \delta_p) - \frac12  n_j(\fvec R + \fvec \delta_p) n_j(\fvec R + \fvec \delta_{p+q})
\end{align}

\section{Ground State at $1/8$ filling}
In this section, we study the ground state at $1/8$ filling, ie.~one particle/hole per unit cell. We cannot solve the ground state analytically with this filling, even in the strong coupling limit. Therefore, we follow the method in the previous section by treating  $\alpha_1$ as a small expansion parameter.

With the ``cluster Hubbard'' terms only, the energy is minimized if each hexagon contains three fermions. As argued in the previous section, both the linear order and the second order of the cross terms vanish. Up to the second order $\mathcal{O}(\alpha_1^2)$, the only contribution to the energy correction comes from
\begin{align}
  & \sum_{\fvec R} \sum_{p} \sum_{q = \pm 1} \sum_{j, \sigma} \sum_{j', \sigma'}  \langle \Psi_{GS} | d^{\dagger}_{j,\sigma}(\fvec R + \fvec \delta_p) d_{j,\sigma}(\fvec R + \fvec \delta_{p + q}) d^{\dagger}_{j',\sigma'}(\fvec R + \fvec \delta_{p + q}) d_{j',\sigma'}(\fvec R + \fvec \delta_{p}) | \Psi_{GS} \rangle \nonumber \\
  = & \sum_{\fvec R, p} \sum_{q = \pm 1} \left\Vert \sum_{j, \sigma} \left. d^{\dagger}_{j,\sigma}(\fvec R + \fvec \delta_p) d_{j,\sigma}(\fvec R + \fvec \delta_{p + q}) \big| \Psi_{GS} \right\rangle \right\Vert^2  \ ,
\end{align}
where $| \Psi_{GS} \rangle$ is the ground state of the ``cluster Hubbard'' terms with the filling of $1/8$.

Similar to the state with $1/4$ filling, we place fermions with the same state (valley and spin) on the honeycomb site. However, half of the sites must be empty for the $1/8$ filling. Therefore, we need to minimize the number of ``dangling'' bonds connecting an occupied site and an unoccupied site. In the stripe phase, each occupied site has exactly one dangling bond, as illustrated in Fig.~\ref{Fig:oneEigth}. For any occupied site, if three neighboring sites are occupied, each of these sites must have two dangling bonds to satisfy the constraint that each hexagon has three fermions. As a consequence, the average of dangling bonds for this ``star'' configuration is $3/2$. Therefore, the number of dangling bonds is minimized in the stripe phase, which is the ground state at least for small $\alpha_1$.

\end{document}